\documentclass[12pt,letterpaper]{article}
\pdfoutput=1
\usepackage{putex}
\usepackage{setspace}
\usepackage{booktabs}
\usepackage{cancel}
\usepackage{multirow}
\usepackage{comment}
\usepackage{bbold}
\usepackage{caption}
\usepackage{amsmath}
\usepackage{enumerate}
\usepackage{cite}
\usepackage{tensor}
\usepackage{slashed}
\usepackage[utf8]{inputenc}
\usepackage{rotating}
\usepackage{bigfoot}

\newcommand{\eq}{\begin{equation}}
\newcommand{\feq}{\end{equation}}
\newcommand{\be}{\begin{equation}}
\newcommand{\ee}{\end{equation}}

\newcommand{\ma}[1]{\mbox{$\mathcal{#1}$}}

\begin{document}

	\begin{center}        
		\Huge Searching for Coleman-de Luccia bubbles in AdS compactifications
	\end{center}
	
	\vspace{0.7cm}
	\begin{center}        
		{\large  Giuseppe Dibitetto$^1$, Nicol\`o Petri$^2$ }
	\end{center}
	
	\vspace{0.15cm}
	\begin{center}  
	\emph{${}^1$ Dipartimento di Fisica, Università di Roma “Tor Vergata” and \\
	Sezione INFN Roma 2, Via della
ricerca scientifica 1, 00133, Roma, Italy.}\\[.2cm]
		\emph{${}^2$ Department of Physics, Ben-Gurion University of the Negev, Be'er-Sheva 84105, Israel.}\\[.3cm]
		\emph{}\\[.2cm]
		e-mails:  \tt giuseppe.dibitetto@roma2.infn.it, \;petri@post.bgu.ac.il
	\end{center}
	
	\vspace{1cm}
	
	
	\begin{abstract}
	\noindent  
	Coleman-de Luccia transitions are spontaneous processes of nucleation of bubbles within metastable gravitational vacua including in their interior a true stable vacuum. From the perspective of lower-dimensional gauged supergravities obtained by truncating Type II and M-theory, these instantonic processes are represented by smooth domain walls featured by de Sitter foliations. These geometries must connect two different AdS vacua in such a way that the wall is defined by an interior and an exterior. We propose a first-order formulation for such radial flows and present two fully backreacted examples of gravitational instantons obtained through this technique, beyond the thin-wall approximation. In the first we consider minimal 7d supergravity describing the truncation of M-theory over a squashed 4-sphere and admitting two AdS$_7$ vacua, one supersymmetric and the other not. Secondly we apply the same strategy to 6d Romans supergravity obtained with consistent truncation of massive IIA supergravity. Also in this case we derive a dS domain wall interpolating between the Brandhuber-Oz vacuum and the non-supersymmetric AdS$_6$ vacuum of the theory.
	\end{abstract}
	
	\thispagestyle{empty}
	\clearpage
	
	\tableofcontents
	
	\setcounter{page}{1}

\section{Introduction}

Ever since the very birth of string theory as a candidate UV complete description of gravitational interactions, it has passed a number of non-trivial theoretical tests that provide a compelling evidence for its UV consistency. These include crucial facts such as black hole microstate counting \cite{Strominger:1996sh}, as well as the AdS/CFT correspondence \cite{Maldacena:1997re}. Moreover, a few preliminary investigations carried out in some controlled setup's even seem to suggest that string theory might actually comprize the entirety of all field theory constructions which are consistent in the UV. This statement is usually referred to as the string universality principle \cite{Kumar:2009us,Adams:2010zy}.

However, all of the above features seem to crucially rely on supersymmetry as a protection mechanism, or at least one could say that we have only been able to draw concrete conclusions within supersymmetric settings. This may be due to the fact that the presence of supersymmetry often allows for analytic treatment and plays a crucial role in making things calculable. In this context, it should not sound surprising that the biggest challenge posed by the string theory paradigm is that of providing a satisfactory mechanism for spontaneous supersymmetry breaking. Needless to say that this is of utmost importance when it comes to connecting our UV consistent theory with a low energy effective model possessing the desired phenomenological properties to describe the world we observe.

In the last two decades, the string universality principle has been addressed by adopting a complementary bottom-up approach, \emph{i.e.} by trying to assess which seemingly consistent field theoretical constructions can actually be UV completetd to fully consistent quantum theories. This way, one might think of coming up with a set of consistency criteria that a low energy description must comply with, in order for it to be related to string theory in an appropriate limit. This approach, \emph{a.k.a.} the string swampland programme \cite{Vafa:2005ui}, has been developing in the last few years in particular and has delivered lots of interesting connections among different desirable IR properties of a given model. We refer the reader to \cite{Palti:2019pca,vanBeest:2021lhn} for a nice review of the recent developments in this field.

The prototypical string swampland conjecture appeared back in the mid 2000's \cite{Arkani-Hamed:2006emk}, where it is argued that a sensible quantum theory describing the interaction between gravity and gauge fields should always retain gravity as the weakest force in the game. The practical criterion proposed there is that there should always exist a microscopic particle in the spectrum, \emph{i.e.} a particle whose mass is smaller than its charge in Planck units.
The authors of \cite{Arkani-Hamed:2006emk} are able to relate this to the common lore that continuous global symmetries should not exist in quantum gravity. 

It is worth noting that, contrary to microscopic particles predicted by the weak gravity conjecture (WGC), macroscopic particles always satisfy a BPS bound instead, \emph{i.e.} $M\geq Q \, M_{\textrm{Pl}}$. In particular, BPS states happen to saturate this inequality, thus being in some sense microscopic and macroscopic objects at the same time. Indeed, the presence of BPS states in the spectrum is protected by supersymmetry all the way from weak to strong coupling. This feature is at the basis of the stability arguments for supersymmetric configurations \cite{Witten:1982df}.

More recently in \cite{Ooguri:2016pdq}, a stronger version of the WGC was proposed, according to which only BPS particles can saturate the BPS bound, this implying that in the absence of supersymmetry there must exist microscopic objects with mass strictly smaller than their charge. Note that, in the context of higher dimensional gravitational theories, the aforementioned objects need not even be particles. These might be extended charged membranes, and their mass should be replaced by the tension. The authors of \cite{Ooguri:2016pdq} infer the non-perturbative instability of non-supersymmetric AdS vacua in string theory as a direct implication of this. Such a non-supersymmetric $d$-dimensional vacuum would then be destroyed by spontaneous nucleation of charged microscopic $(d-2)$ membranes that eventually discharge the flux supporting the original vacuum. 

It is worth stressing that this instability may result in the absence of a holographic CFT dual. From a gravitational viewpoint, this may be seen as a consequence of the fact that any local instability occuring somewhere in the bulk would take a finite global time to reach the boundary of AdS, where its dual CFT is supposed to live \cite{Harlow:2010az}.  If one has the underlying brane picture in mind, what prevents taking the conformal limit in a non-supersymmetric setting is the impossibility to pile up the branes into a stack, as they would feel a force repelling each other due to the WGC. As a consequence, checking non-perturbative stability of AdS vacua is crucial for their holographic interpretation to hold. However, this is in general not easy a task within the full stringy description of the given vacuum. We refer to \cite{Narayan:2010em} as an example where the proper stringy treatment has been illustrated.

On the other hand, non-perurbative decay processes such as gravitational tunneling events have been widely studied since the beginning of the 1980's \cite{Coleman:1980aw, Brown:1988kg} within the framework of semiclassical Euclidean path intergrals in 4d gravity. The tunneling process between two classical vacua is seen as a bubble nucleation event for observers living in the false vacuum geometry. Enclosed within the bubble wall, there is the true vacuum geometry. Viable decay channels correspond to local extrema of the Euclidean action. In the thin shell limit, these are identified by junctions respecting the so-called Coleman-de Luccia (CDL) bound \cite{Coleman:1980aw}. Going beyond the thin limit means being able to dynamically source the jumps in the fields across the aforementioned junction by employing extra scalar fields. This procedure generically results in a smooth interpolating geometry called domain wall (DW). 

DW solutions have been extensively explored in the context of lower dimensional supergravity theories. In particular, when supersymmetry is preserved, these turn out to obey first order flow equations determined by the corresponding Killing spinors. This often allows one to analytically determine the corresponding dynamical profiles. In this paper we will show how positively curved DW's are directly related to the aforementioned CDL bubbles and are intrinsically non-supersymmetric. Despite this, we will make use of the Hamilton-Jacobi (HJ) formalism in order to provide a first-order formulation thereof \cite{Skenderis:2006rr}. This will allow us to overcome the problem of sensitivity to the choice of initial data, which the second-order formulation suffers from.

Similar techinques were recently used in \cite{Ghosh:2021lua}, where positively curved domain wall solutions connecting different AdS vacua were explicitly found. The setup there is a generic Einstein-scalar theory with the addition of a negative cc, though the explicit choices of scalar potentials are not directly linked to a specific higher dimensional origin. Besides discussing in detail their interpretation on the gravity side as CDL bubbles, the authors of \cite{Ghosh:2021lua} also identify their holographic description to be a CFT on a cylinder $\mathbb{R}\times S^{d-1}$, rather than on flat space.

In our work, we want to take a further step and connect our lower-dimensional solutions to string and M-theory. The crucial ingredient to achieve this will be the existence of consistent truncations \cite{Cvetic:2000dm}. To this end, we will apply the aforementioned technique to lower-dimensional supergravity models in dimension six and seven, which are known to arise from consistent truncations of massive type IIA supergravity \cite{Cvetic:1999un} and 11d supergravity \cite{Lu:1999bc} on a (squashed) four-sphere, respectively. At this point, any lower-dimensional solution (including our CDL bubbles) have a natural higher-dimensional interpretation. A much harder task would then be that of understanding these bubbles as composite objects made out of fundamental stringy or M-theoretical building blocks, such as strings and membranes. For the moment, we will leave this out for future investigation.

The paper is organized as follows. In section \ref{Sec:CDL}, we briefly review the basics of CDL bubbles and their relation to the general classification of DW's by \cite{Cvetic:1992jk}. In section \ref{bubbles7d} we review the consistent truncation of 11d supergravity on a squashed $S^4$. Subsequently, within the truncated 7d supergravity theory, we present the numerical solution of interest describing a CDL instanton within the non-supersymmetric AdS$_7$ vacuum of the theory. In section \ref{bubbles6d}, we apply the same machinery within Romans' supergravity in 6d, which stems from massive type IIA supergravity on $S^4$. In appendix \ref{HJmethod} we present some general facts and introduce some notation concerning the HJ formalism.

\section{Coleman-de Luccia decays and dS domain walls}
\label{Sec:CDL}

The process of gravitational tunneling can be seen as the spontaneous nucleation of a true vacuum bubble within a false vacuum geometry, and it can be studied within a semiclassical regime by Euclidean path integral techinques. This was done already in the early 1980's \cite{Coleman:1980aw} by adopting the thin-wall approximation, where an infinitely thin bubble wall separates two different regions, each one characterized by a different value of the cosmological constant, say $\Lambda_{\pm}$. In \cite{Coleman:1980aw} it was found that in 4d a critical value for the tension of the bubble wall (expressed in Planck units)
\be\label{CDL_bound}
\sigma \, \overset{!}{\leq} \,\sigma_{\textrm{CDL}} \, \equiv \, \frac{2}{\sqrt{3}}\left(-\sqrt{|\Lambda_{+}|} \, + \, \sqrt{|\Lambda_{-}|}\right) \ ,
\ee
is an upper bound in order for the Euclidean action to admit a local extremum. This is usually referred to as the Coleman-de Luccia (CDL) bound. Later in \cite{Brown:1988kg}, it is shown that the mechanism of extremization for the Euclidean action can be physically understood as imposing energy conservation during the nucleation process. In this context, the CDL bound represents the allowed range of wall tensions admitting a finite bubble radius guaranteeing energy conservation. For a wall tension exactly saturating the CDL bound, the bubble radius becomes infinite, thus yielding a flat static  wall rather than an actual bubble. For values larger than this critical value, no real bubble size turns out to be compatible with energy consevation. Once the Euclidean action is extremized at some finite value, the corresponding instanton configuration contributes to the decay rate of the vacuum through
\be
\Gamma_{\textrm{decay}} \, \sim \, e^{-S_{E}(\textrm{instanton})} \ .
\ee

Back in Lorentzian signature, these instanton geometries correspond to DW's connecting two different vacuum solutions. Within the thin-wall approximation, these DW's are obtained by gluing two different maximally symmetric vacuum solutions with different values of the cosmological constant to each other, across a certain interface. Such DW's were studied and classified in \cite{Cvetic:1992jk}. The general form of the metric reads
\be
ds^{2}_{d+1} \, = \, dr^{2} \, + \, e^{2A(r)}L^{2}ds_{d}^{2} \ ,
\ee
where the metric for the $d$ dimensional slices $ds_{d}^{2}$ is chosen to be maximally symmetric, and the asymptotic behavior of the warp factor $A$ must be specified in such a way that the asymptotic geometries on the two sides of the wall are $(d+1)$ dimensional maximally symmetric vacua\footnote{Note that the constant $L$ just represents some reference length that makes $ds_{d}^{2}$ dimensionless. For (A)dS it just represents the (A)dS radius, while in Mkw it is completely irrelevant, as it can reabsorbed into a redefinition of the spacetime coordinates.}. Let us now explore in detail all the possible features that fully characterize DW solutions. 

\subsection*{Domain wall zoology}
Once denoted by $\Lambda_{\pm}$ the cosmological constants on each side, and by $\kappa$ the curvature parameter of the wall, solving the Einstein equations in the thin wall limit turns out to be equivalent to solving the so-called Israel junction condition \cite{Israel:1966rt}. This essentially fixes the DW tension as to compensate for the jump in extrinsic curvature across the wall. As a result, the DW tension in Planck units\footnote{We also drop here $\mathcal{O}(1)$ constant factors that are not crucial for the sake of our conceptual treatment. Besides, these would make it harder to provide a general discussion in arbitrary $d$, as they explicitly depend on the number of spacetime dimensions.} can be expressed as \cite{Cvetic:1992jk}
\be\label{DW_tensions}
\sigma \, = \, \eta_{+}\sqrt{\frac{\kappa}{L^2}-\Lambda_{+}} \, + \, \eta_{-}\sqrt{\frac{\kappa}{L^2}-\Lambda_{-}} \ ,
\ee
where $\eta_{\pm}\in\{\pm1\}$ represent the orientations of the normal vector on the two sides of the wall. In particular, $-1$ indicates an exterior, while $+1$ indicates an interior.

In order then to fully characterize a DW, it turns out to be sufficient to specify its thin wall data, \emph{i.e.}
\begin{itemize}
\item the signs of $\Lambda_{\pm}$ ($0$ allowed),
\item the DW curvature parameter $\kappa=0,\pm1$,
\item the signs of $\eta_{\pm}$.
\end{itemize}
The first choice selects the case of interest out of a list of six possibilities \cite{Cvetic:1992jk}: dS/dS, dS/Mkw, dS/AdS, Mkw/Mkw, Mkw/AdS, AdS/AdS. It is worth noting that the first three involving dS space never admit the extreme limit, \emph{i.e.} taking $L\rightarrow\infty$. This applies to some extent to Mkw/Mkw DW's as well, as extremal ones have vanishing tension. The remaining two choices, which do admit an extreme limit, were studied in detail in \cite{Cvetic:1992bf} in the supersymmetric BPS case, where extremality is guaranteed. There the analysis is performed even beyond thin wall, thanks to first order flow equations implied by the existence of Killing spinors.

Of particular interest to our scope, is the case of AdS/AdS DW's, since we will see that, under certain circumstances, they will precisely describe CDL bubbles. In \cite{Cvetic:1992bf}, the key feature distinguishing different AdS/AdS DW's is whether or not the superpotential $W$ vanishes somewhere along the flow. This identifies a non-monotonic flow (dubbed type II), or rather a monotic one (dubbed type III), respectively. In the thin wall limit this is directly related to the choice of orientation ($\eta_{\pm}$) on the two sides, determining if the wall has two insides (I/I), two outsides (O/O), or an inside and an outside (I/O).

If we now compare the CDL critical tension in \eqref{CDL_bound} with \eqref{DW_tensions}, we immediately see that they exactly coincide for $\kappa=0$ and $\eta_{+}=-1$, $\eta_{-}=+1$, \emph{i.e.} an extremal (flat) I/O DW. For I/O walls with general $\kappa$, the tension reads
\be
\sigma \,=\, \sqrt{\frac{\kappa}{L^2}-\Lambda_{-}} \, - \, \sqrt{\frac{\kappa}{L^2}-\Lambda_{+}} \ \longrightarrow \ \left\{ \begin{array}{lcclc}
> \ \sigma_{\textrm{CDL}} & , & & \kappa=-1 & , \\
= \ \sigma_{\textrm{CDL}} & , & & \kappa=0 & ,\\ 
< \ \sigma_{\textrm{CDL}} & , & & \kappa=+1 & .\end{array}\right.
\ee
This implies that actual gravitational instantons such as CDL bubbles are represented by I/O DW's with dS slices \cite{Cvetic:1992st}.
\begin{table}[htp]
\begin{center}
\begin{tabular}{|c|c|c|c|}
\hline
$\left(\eta_{-},\eta_{+}\right)$ & Orientation & Tension & Physical Interpretation  \\[2mm]
\hline \hline
$(+1,+1)$ & I/I & $\sigma \ > \ \sigma_{\textrm{CDL}}$ & Wormhole \\
\hline
$(+1,-1)$ & I/O & $0 \ < \ \sigma \ < \ \sigma_{\textrm{CDL}}$ & CDL bubble \\
\hline
$(-1,-1)$ & O/O & $\sigma \ < \ 0$ & RS brane \\
\hline
\end{tabular}
\end{center}
\caption{\it The inequivalent AdS/AdS spherical ($\kappa=+1$) domain walls. CDL bubbles are positively curved DW's with an inside and an outside. The remaining two situations, corresponding to exotic objects such as wormholes and RS branes, do not create spontaneously.}
\label{tab:DW_tensions}
\end{table}

On the other hand, for a $(-1,-1)$ choice of orientation (corresponding with a wall with two outsides), one has an object with negative tension. The effect of placing this wall at some finite $r$ in DW coordinates is that of cutting out a portion of infinite volume from spacetime, thus rendering the total effective volume finite. This is exactly a Randall-Sundrum (RS) brane, which was used in \cite{Randall:1999vf} in order to produce an effective 4d braneworld within AdS$_5$ in such a way that 4d gravity be localized on the brane. While locally such a spacetime looks like AdS on both sides, globally it has no boundaries, since they have been removed on both sides. 

Finally, if one considers the situation with two insides, the tension overcomes the CDL bound. This means that such an object would not be spontaneously created, but it is rather an exotic extended source that formally solves the junction condition. The local geometric structure still looks like AdS on both sides, but globally, unlike the RS brane, it has two boundaries placed at finite proper distance. In this sense, we qualify this object as a wormhole.
The physically inequivalent AdS/AdS positively curved DW's are illustrated in table \ref{tab:DW_tensions}.

In the following sections, we will consider two examples of lower-dimensional supergravity theories with a known higher-dimensional origin and we will show how to construct I/O AdS/AdS positively curved DW's. As we have just been arguing, these correspond with CDL bubbles describing non-perturbative decays of an AdS false vacuum into an AdS true one. The technique we will make use of is the HJ formalism, which allows us to recast the original second-order differential problem into a first-order one. Thanks to this, we will get rid of the fine-tuning problem related to the choice of initial data, as it will be completely hidden into the correct choice of HJ generating functional. Upon numerical integration, we will obtain the desired solutions.

\section{Bubble geometries in M-theory}\label{bubbles7d}

In this section we derive a smooth bubble geometry connecting two different AdS$_7$ vacua, one supersymmetric and the other not. Our framework will be $\ma N=1$ 7d supergravity obtained by truncating M-theory on a squashed 4-sphere using the compactification formulas of \cite{Lu:1999bc}. Our aim is to consider domain walls driven by one single scalar field and featured by a dS$_6$ slicing. 

For such geometries we construct the quantities needed in order to cast the second-order problem of the equations of motion into a first-order one, supplemented with one single extra PDE, the Hamilton-Jacobi equation. The HJ formulation of classical systems, summarized in Appendix \ref{HJmethod}, allows us to formulate the problem of finding the dynamics of the aforementioned domain walls in terms of a set of first-order ODEs whose solutions turn out to authomatically satisfy the field equations.

The key of this procedure is finding the fake superpotential (the HJ generating functional) solving the Hamilton-Jacobi equation and defining the first-order system. In this section we discuss various strategies to solve this problem for domain walls interpolating between the two AdS$_7$ vacua of the theory. Finally we obtain an explicit numerical solution for this fake superpotential using a perturbative method whose efficiency perfectly suits this particular situation. Once obtained the solution for the fake superpotential, we derive the radial flow of the interpolating domain wall.

\subsection{Supergravity setup and AdS$_7$ vacua}\label{7dtheory}

It is well-known that M-theory can be consistently reduced over 4-spheres (see \emph{e.g.} \cite{Pilch:1984xy,Nastase:1999cb,Lu:1999bc}). In this section we are interested in the minimal truncation of 11d supergravity, namely the truncation retaining only the 7d fields belonging to the supergravity multiplet (\emph{i.e.} no matter couplings). The compactification reproducing such a theory is a warped compactification over a squased $S^4$ and it has been worked out in \cite{Lu:1999bc}. Apart from the 7d gravitational field, this dimensional reduction yields one real scalar field $X$, three SU$(2)$ vectors $A^i$ and a 3-form $B_{(3)}$ \cite{Townsend:1983kk}.

For the aim of this paper we can focus on the case where the vectors and the 3-form are vanishing. The 11d truncation Ansatz takes then the simplified form \cite{Lu:1999bc}
\begin{equation}
 \begin{split}\label{truncationmetric7d}
  ds^2_{11}&=\Delta^{1/3}\,ds^2_7+2g^{-2}\,\Delta^{-2/3}\,ds^2_{4} \ ,\\
  ds_{4}^2&=\Delta\,X^3\,d\xi^2+X^{-1}\,c^{2}\,ds^2_{S^3}\ ,\qquad \text{with} \qquad \Delta=X\,c^2+X^{-4}\,s^2\ ,
 \end{split}
\end{equation}
where, for simplicty of notation, $s=\sin \xi$ and $c=\cos\xi$. With the assumption of vanishing 7d form fields, the 11d 4-form boils down to \cite{Lu:1999bc}
\begin{equation}
\begin{split}\label{truncationfluxes7d}
  G_{(4)}&=-2\sqrt 2\,g^{-3}\,c^{3}\,\Delta^{-2}\,\left(X^{-8}\,s^2-2X^2\,c^2+3\,X^{-3}\,c^2-4\,X^{-3}\right)\,d\xi\,\wedge\,\text{vol}_{S^3}\\
  &-10\sqrt{2}\,g^{-3}\,\Delta^{-2}\,X^{-4}\,c^4\,s\,dX\,\wedge\,\text{vol}_{S^3}.
 \end{split}
\end{equation}
The above truncation defines a gauged supergravity with $\mathbb{R}^+\times \mathrm{SO}(3)$ symmetry and featured by two gauge couplings $g$ and $h$. The first is associated to the R-symmetry $\mathrm{SU}(2)_R$ group that is gauged in this theory and the second is a St\"uckelberg mass for the 3-form. The scalar potential for the scalar $X$ has the form
\begin{equation}\label{pot7d}
 V=2h^2\,X^{-8}-16\,h^2\,X^{-3}-16\,h^2\,X^2\,,
\end{equation}
where the explicit truncation of \cite{Lu:1999bc} has fixed $g=2\sqrt 2\,h$. The Lagrangian of this theory is thus given by
\begin{equation}
\label{action7d}
 \sqrt{-g}^{\,-1}\ma L=  R  -5\, X^{-2}\, \partial_\mu X\,\partial^\mu X-V\,,
\end{equation}
leading to the equations of motion
\begin{equation}
 \begin{split}\label{eom7d}
  R_{\mu\nu}-5\,X^{-2}\,\partial_\mu X \partial_\nu X-\frac 15 \, V\, g_{\mu\nu}&=0\ ,\\
  \partial_\mu \left(\sqrt{-g}\,X^{-1}\,g^{\mu\nu}\,\partial_\nu X\right)-\frac{\sqrt{-g}}{10}\,X\,\partial_X V&=0\ .
 \end{split}
\end{equation}
This theory has two AdS$_7$ vacua, a supersymmetric and a non-supersymmetric one. Let us now consider them separately.

\vspace{.5cm} {\bf SUSY AdS$_7$ vacuum: $X=1$}

\noindent This vacuum is preserving 16 real supercharges and it is realized for $X=1$. The 11d background takes the form of a direct product of AdS$_7$ and a round $S^4$,
\begin{equation}
\begin{split}\label{SUSYAdS7}
 &ds^2_{11}=ds^2_{\text{AdS}_7}+2g^{-2}\,ds^2_{S^4}\,,\\
  & G_{(4)}=6\sqrt{2}\,g^{-3}\,c^3\,d\xi \wedge\text{vol}_{S^3}\,,
 \end{split}
\end{equation}
with the radius of AdS$_7$ given by $L_{\text{SUSY}}=2\sqrt 2\,g^{-1}=h^{-1}$.
The brane interpretation of the above solution is clear. In fact it can be viewed as the Freund-Rubin vacuum associated with the near-horizon geometry of a stack of M5 branes\footnote{This maximally supersymmetric interpretation appears by considering this solution within the maximal $\mathrm{SO}(5)$ gauged theory. There a SUSY enhancement to $32$ supercharges becomes manifest.}. Alternatively one can also view \eqref{SUSYAdS7} as the half-supersymmetric 11d vacuum arising from M5 branes on an A-type singularity.

\vspace{.5cm} {\bf Non-SUSY AdS$_7$ vacuum: $X=2^{-1/5}$}

\noindent A non-supersymmetric AdS$_7$ vacuum can be obtained by setting $X=2^{-1/5}$. In this case the geometry is warped
\begin{equation}
\begin{split}\label{nonSUSYAdS7}
 &ds^2_{11}=2^{-1/15}\left(2-c^2\right)^{1/3}\left[ds^2_{\text{AdS}_7}+2^{2/5}g^{-2}\,d\xi^2+ 2^{7/5}g^{-2}\,\left(2-c^2\right)^{-1}c^2\,ds^2_{S^3}\right],\\
  & G_{(4)}=8\sqrt 2\,g^{-3}\,c^3\left(2-c^2\right)^{-2}\,d\xi \wedge\text{vol}_{S^3}\,,
 \end{split}
\end{equation}
with the radius of AdS$_7$ given by $L_{\slashed{\text{SUSY}}}=2^{7/10}\sqrt 3\,g^{-1}=\frac{\sqrt 3}{2^{4/5}}\,h^{-1}$.
Unlike the supersymmetric case, the brane interpretation of this vacuum is not clear as well as its derivation as near-horizon limit of some 11d brane solution.

\subsection{First-order formulation for dS$_6$ domain walls}\label{HJ7d}

Let us now take the 7d theory \eqref{action7d} as operative framework and study domain walls of the following type
\begin{equation}
 \begin{split}\label{dSDW}
  ds^2_{7}&=e^{2A(r)}\,L^2\,ds^2_{\text{dS}_6}+dr^2\ ,\\
  X&=X(r)\ .
 \end{split}
\end{equation}
As it is manifest the worldvolume of these domain walls is curved by a dS$_6$ space described by the element $ds^2_{\text{dS}_6}$ with radius $L$. The two functions $A(r)$ and $X(r)$ specify the (11d) geometry. This class of backgrounds is what we need, in order to study an expanding bubble within a given vacuum. Of course also some global aspects are crucial in order to describe gravitational instantons and the false vacuum decay \cite{Cvetic:1993xe}. At the current stage of analysis, we may say that the search of such domain walls connected to AdS$_7$ vacua, implies a particular choice of boundary conditions on the vacuum geometry at infinity, namely foliating the vacua with dS. This parametrization is in fact the one in which the topology of an expanding bubble is manifest.
The field equations \eqref{eom7d} split into two second-order non-linear ODEs for $A$ and $X$,
\begin{equation}
 \begin{split}\label{eomDW}
   A''+6A'^{\,2}-\frac{5\,e^{-2A}}{L^2}+\frac{1}{5}\, V&=0\ ,\\
    X''+6 A'\,X'-\frac{X'^{\,2}}{X^{2}}-\frac{1}{10}\,X^2\, \partial_X\,V&=0\ ,\\
 \end{split}
\end{equation}
and a first-order differential constraint
\begin{equation}\label{Hconstraint}
30A'^{\,2}-\frac{5\,X'^{\,2}}{X^{2}}-\frac{30\,e^{-2A}}{L^2}+V=0\ ,
\end{equation}
where $'$ denotes the derivative with respect to $r$.
The condition \eqref{Hconstraint} is the analogue of the first Friedman equation (for real time cosmology) and is usually called ``Hamiltonian constraint'' in this context. This condition must be dynamically satisfied by any solution of \eqref{eomDW}.
The vacuum geometry can be obtained by choosing
\begin{equation}\label{Avacua}
 A=\log \left(\sinh\left(\frac{r}{L} \right)\right)\,,
\end{equation}
and imposing $X=1$ and $L_{\text{SUSY}}=h^{-1}$ or $X=2^{-1/5}$ and $L_{\slashed{\text{SUSY}}}=\frac{\sqrt 3}{2^{4/5}\,h}$ respectively for SUSY and non-SUSY cases discussed in Section \ref{7dtheory}.

Following the general discussion of Appendix \ref{HJmethod}, we can look at the second-order problem \eqref{eomDW} and \eqref{Hconstraint} as a classical constrained system with dynamical variables $A$ and $X$. Our aim is to recast the second-order problem into a first-order one by using the standard Hamilton-Jacobi formulation of classical dynamics. First of all we point out that the 2nd order equations \eqref{eomDW} can be obtained from the effective Lagrangian,
\begin{equation}\label{effectiveL}
 L_{\text{eff}}=30\,e^{6A}\,A'^{\,2}-5\,e^{6A}\,\frac{\,X'^{2}}{X^{2}}+\frac{30\,e^{4A}}{L^2}-e^{6A}\,V\,.
\end{equation}
From this expression we can extract the corresponding Hamiltonian by introducing the conjugate momenta
$\pi_A=60\,e^{6A}\,A'$ and $\pi_X=-10\,e^{6A}\,X^{-2}\,X'$ following the general expression \eqref{classicalH}. Taking the Legendre transformation of \eqref{effectiveL} leads to the Hamiltonian
\begin{equation}
 H_{\text{eff}}=\frac{1}{120}\,e^{-6A}\,\pi_A^{2}-\frac{1}{20}\,e^{-6A}\,X^{2}\,\pi_X^{\,2}-\frac{30\,e^{4A}}{L^2}+e^{6A}\,V\,.
\end{equation}
We can now introduce the fake superpotential $F=F(A,X)$ associated with the system \eqref{eomDW} as the real function such that $\pi_A=\partial_A\,F$ and $\pi_X=\partial_X\,F$ satisfying the Hamilton-Jacobi equation
\begin{equation}\label{HJequation}
 \frac{1}{120}\,e^{-6A}\,(\partial_A\,F)^{2}-\frac{1}{20}\,e^{-6A}\,X^{2}\,(\partial_X\,F)^{2}-\frac{30\,e^{4A}}{L^2}+e^{6A}\,V=0\,,
\end{equation}
where the constant $E$ appearing in the general expression \eqref{genericHJequation} must be zero in order to satisfy the Hamiltonian constraint \eqref{Hconstraint}. We thus reduced the constrained the system of ODEs \eqref{eomDW}, \eqref{Hconstraint} to a single PDE that have to be satified by a suitable solution for the superpotential $F$. The radial dependence of the functions describing the domain walls \eqref{dSDW} can be thus obtained by specifying the first-order constraints \eqref{general1STorder} to our particular case, namely
\begin{equation}\label{1storderflow}
  A'=\frac{1}{60}\,e^{-6 A}\partial_A\,F\qquad \text{and} \qquad X'=-\frac{1}{10}\,e^{-6 A}X^2\,\partial_X\,F\,.
\end{equation}
As explained in Appendix \ref{HJmethod} the solutions of the above 1st order equations automatically solve also the equations of motion, since imposing the conditions \eqref{1storderflow} is equivalent to extremizing the action \eqref{effectiveL}.

\subsection{Strategies of integration}\label{strategies}

Let us consider domain wall geometries of type \eqref{dSDW} interpolating between two different AdS$_7$ vacua \eqref{Avacua}, namely smooth solutions that, on one side, asymptotically reproduce the supersymmetric vacuum with $X=1$ and on the other side the non-supersymmetric one with $X=2^{-1/5}$.
In the last section we showed how to cast the 2nd order problem of the equations of motion \eqref{eomDW} in terms of a single PDE \eqref{HJequation} for a superpotential $F(A,X)$. We want now to discuss possible strategies in solving the dynamics for this particular domain wall geometry.

The presence of the de Sitter foliation and our particular requirement on the asymptotic behavior make the search for explicit solutions for the superpotential a hard task. In fact the contribution to the stress-energy tensor associated to the curvature of the dS$_6$ foliation does not allow to formulate a separable Ansatz for $F(A,X)$ and this forces us to approach the problem with numerical methods. This can be clearly seen by looking at the Hamilton-Jacobi equation \eqref{HJequation},
\begin{equation}\label{HJequation1}
 \frac{1}{120}\,e^{-6A}\,(\partial_A\,F)^{2}-\frac{1}{20}\,e^{-6A}\,X^{2}\,(\partial_X\,F)^{2}+V_{\mathrm{eff}}=0\ ,
\end{equation}
where we introduced the ``effective" potential,
\begin{equation}\label{Veff}
 V_{\mathrm{eff}}=-\frac{30\,e^{4A}}{L^2}+e^{6A}\,V\ .
\end{equation} 
\begin{figure}[!http]
	\centering
	\includegraphics[width=0.75\textwidth]{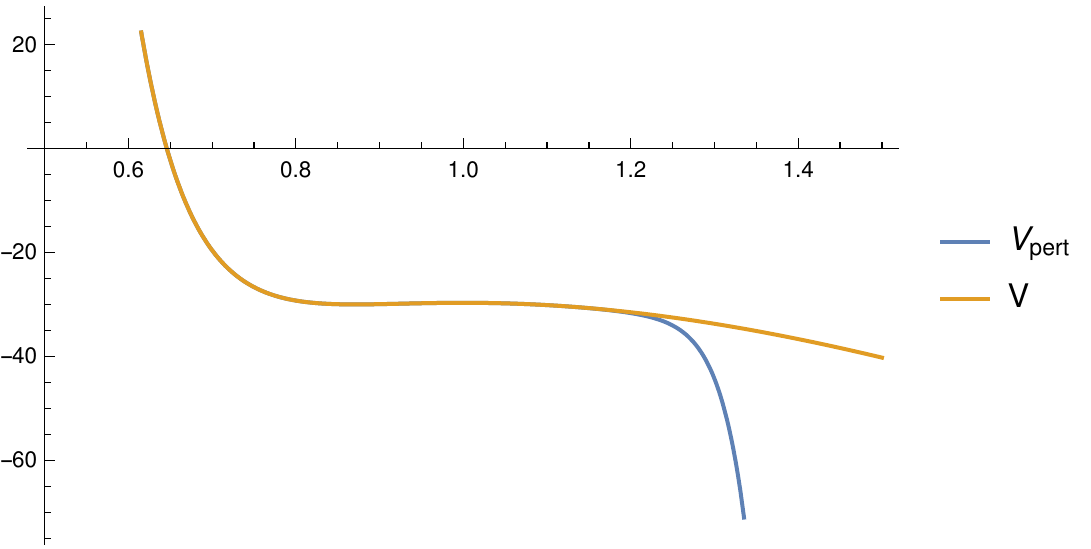}
	\caption{\it  The comparison between the scalar potential $V(X)$ written in \eqref{pot7d} and its expansion at 15th order around the non-SUSY critical point $X=2^{-1/5}\simeq 0.87$. For values of $X$ between the two critical points $X\,\in\,[2^{-1/5},\,1]$, the scalar potential is well-approximated by its expansion. The plot is given for $h=2\,\sqrt{2}\,g=1$.
	}
	\label{fig:comparePotentials}
\end{figure}
The first term in \eqref{Veff} is associated to the curvature of the dS$_6$ slicing and it does not allow to separate the variables\footnote{In the flat limit $L\rightarrow \infty$ the dS$_6$ foliation can be substituted by 6d Minkowski spacetime. In this case the superpotential takes the separable form $F=4 e^{6A}\,f(X)$ with $f=-\frac h2 \left(X^{-4}+4X \right)$ such that $V=-\frac 45\, \left(6f^2-X^2\partial_X\,f^2 \right)$.}.

In addition to this, the requirement of having different AdS$_7$ vacua in the two asymptotic regions does not mean that we can treat them as equivalent initial conditions for our integration. This can be seen by trying to perform a direct numerical integration of 2nd order equations of motion \eqref{eomDW} by starting from a linearized expansion of the supersymmetric vacuum. Such integrations generically produce singular behaviors (connected with M5 sources) at the other end of the flow. From these attempts it is manifest that the particular solution for $A$ and $X$ converging to the non-SUSY vacuum is determined by a set of initial parameters constituting a null measure set within the parameter space describing the linearized expansion around the SUSY vacuum. This can be rephrased by observing that the SUSY vacuum plays the role of  an attractor point in the space of solutions (just like an M5 singularity does!), while the non-SUSY one does not.

Given all of the above, we are lead to approach the problem of integrating the Hamilton-Jacobi equation \eqref{HJequation1} by starting from a perturbative expansion of the fake superpotential $F(A,X)$ around the non-SUSY vacuum, namely the point $X=2^{-1/5}$ in the moduli space. The idea is to apply the perturbative method discussed in \cite{Dibitetto:2022vdw}\footnote{The perturbative method discussed in this reference further extends the one presented in \cite{Danielsson:2016rmq}.} to this particular situation and to solve order-by-order the perturbative tower of ODEs reproduced by the Hamilton-Jacobi equation when the superpotential is expanded around the non-SUSY vacuum. As it was pointed out in \cite{Dibitetto:2022vdw}, in the case of domain walls with a curved worldvolume, the coefficients of the aforementioned expansion must depend on the warp factor of the metric $A$ and this constitutes the main complication with respect to the case of domain walls with a flat slicing.

This procedure turns out to perfectly suit the current situation and we can test its effectiveness by comparing the 7d scalar potential $V(X)$ written in \eqref{pot7d} with its Taylor expansion $V_{\text{pert}}(X)$ around the non-SUSY vacuum $X=2^{-1/5}$ as in Figure \ref{fig:comparePotentials}. For values of $X$ included within the two extrema $X\,\in\,[2^{-1/5},\,1]$, at sufficiently high order, the scalar potential nicely coincides with its perturbative expansion, up to a great level of accuracy.

\subsection{The interpolating solution}\label{perturbativeHJ}

Since we are going to solve the Hamilton-Jacobi equation \eqref{HJequation1} with a perturbative expansion of the fake superpotential, we need to specify boundary conditions for $F$ (or more precisely on its derivatives) that can be used to generate the initial conditions for the integration at each order. Such boundary data can be obtained by solving the Hamilton-Jacobi equation \eqref{HJequation1} for small values of the radius $L$. This regime is featured by a dominant contribution of the dS$_6$ curvature in the effective potential \eqref{Veff}. In particular one can verify that the expression
\begin{equation}\label{Finfty}
 F_{\infty}=\frac{12}{L}\,e^{5A}-\frac{L}{7}\,e^{7A}\,V
\end{equation}
solves the equation \eqref{HJequation1} up to $O(L^2)$ as $L \ll \ell_{\mathrm{AdS}}$. As confirmed by the plot of Figure \ref{fig:comparePotentials}, we can thus substitute $V$ by its Taylor expansion $V_{\text{pert}}$ around the point $X=2^{-1/5}$ provided that the expansion be sufficiently large to reproduce a good matching of the curves in the interval $X \in \left [2^{-1/5},1 \right]$. In this way we can produce a set of initial conditions at each order in the expansion.

We can now write a perturbative Ansatz of the form\footnote{The factor of 4 is needed to reproduce the flat limit given by large values of $L$.}
\begin{equation}\label{perturbativeF}
 F(A,X)=4\,\sum_{k=0}^{\infty}F^{(k)}(A)\,\frac{(X-2^{-1/5})^k}{k!}\ ,
\end{equation}
where the coefficients $F^{(k)}$ crucially depend on $A$ \cite{Dibitetto:2022vdw} in order to take into account the non-separability of the effective potential \eqref{Veff}. The Hamilton-Jacobi equation \eqref{HJequation1} thus reproduces at $X=2^{-1/5}$ a set of ODEs for the coefficients $F^{(k)}(A)$, each for any order in $k$. 
First of all we observe that we need to impose that $F^{(1)}(A)=0$ in order to keep each perturbative order decoupled. The 0-th order equation takes the form
\begin{equation}\label{zeroorder}
 \frac{2}{15}\,e^{-6A}\,(\dot F^{(0)})^2-\frac{30\,e^{4A}}{L^2}-\frac{30\,e^{6A}}{L^2}=0\ ,
\end{equation}
\begin{figure}[!http]
\centering
	\includegraphics[scale=0.5]{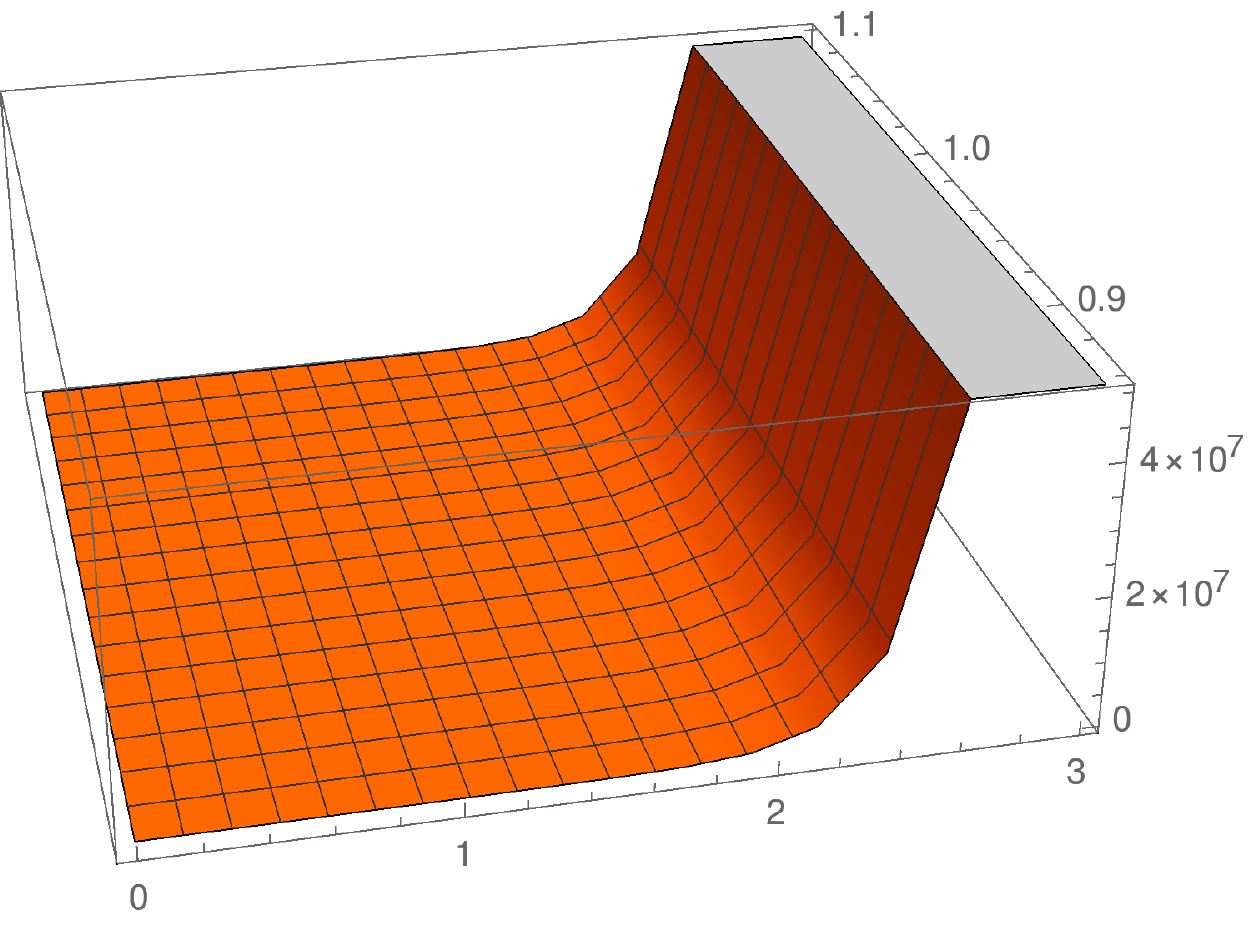}\qquad\quad \,\,\,
	\centering
	\includegraphics[scale=0.6]{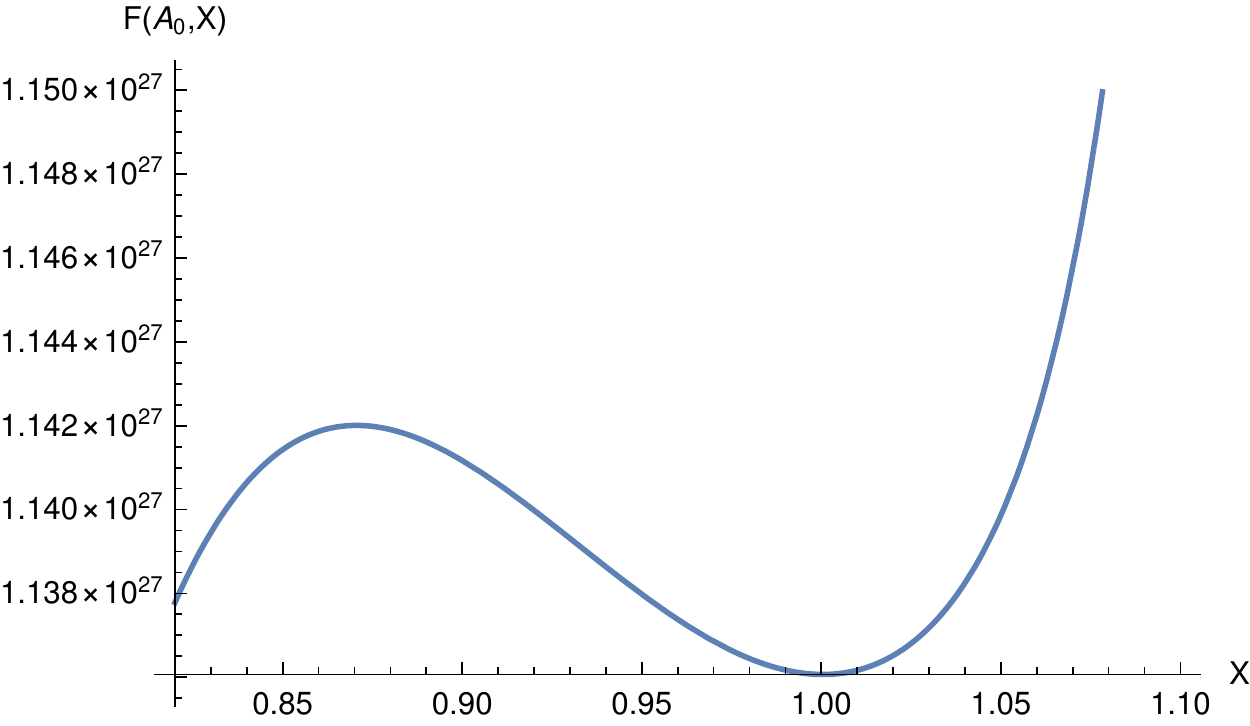}
	\caption{\it The profile of the fake superpotential $F(A,X)$ obtained by iterating the perturbative intergrations up to order $k=15$. The 3d plot describes the full solution in the intervals $A\in[0,\,3]$ and $X\in[0.86,1.1]$. On the right side the same function is plotted for $A_0=10$. Even if the variation of $F$ is tiny, we observe an interpolating profile between the two AdS$_7$ vacua respectively located at $X=2^{-1/5}\simeq0.87$ and $X=1$. The plots are given for $L=1$.
	}
	\label{fig:plotSuperpotential}
\end{figure}
where we denoted with $\cdot$ the derivative with respect to $A$ and we expressed the scalar potential in terms of the radius of non-SUSY vacuum through the relation $h=\frac{2^{4/5}}{\sqrt 3}L$ derived in \eqref{nonSUSYAdS7}. This equation can be solved exactly,
\begin{equation}\label{Fzero}
 F^{(0)}(A)=\frac{5}{16\,L}\, e^A\sqrt{1+e^{2A}}\,\left(-3+2\,e^{2A}+8\,e^{4A}\right)+\frac{15}{16L}\text{arcsinh} \left(e^A \right)\ .
\end{equation}
The subsequent step is to determine $F^{(2)}(A)$ by integrating the ODE appearing at order $k=2$ with $F^{(0)}(A)$ given in \eqref{Fzero}. This is a non-linear ODE and it can be integrated numerically. Once this equation is solved (with a suitable initial condition that we specify below), each perturbative order in $k$ can be solved iteratively in terms of the previous ones, since each of these equations turns out to be linear in $F^{(k)}(A)$. At each order in $k\geq2$ we thus perform a numerical integration evaluating the $(k-1)$-th solution of the previous step in the $k$-th ODE for $F^{(k)}(A)$. As initial conditions we choose
\begin{equation}\label{initF}
 F^{(k)}=4^{-1}\,\partial^k_{X}\,F_{\infty}^{(k)}|_{X=2^{-1/5}}\ ,
\end{equation}
where $F_{\infty}^{(k)}$ is given by \eqref{Finfty} written in terms of the Taylor expansion of the scalar potential $V$ at the $k$-th order.
Iterating the numerical integrations up to order $k=15$ we obtain the solution plotted in Figure \ref{fig:plotSuperpotential}. The relevant behavior can be observed in the plot on the right side where an interpolating behavior of $F$ between the two vacua is manifest. This is the key property allowing the existence of interpolating domain walls.
\begin{figure}[!http]
\centering
	\includegraphics[scale=1]{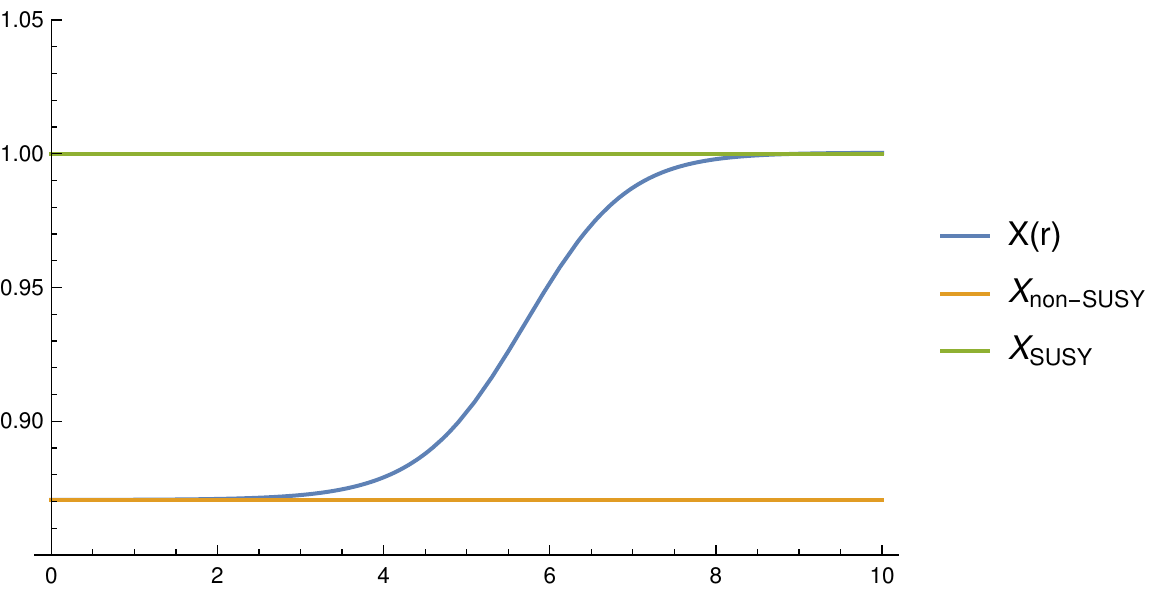}
	\caption{\it The radial profile for the scalar $X$ in the dS$_6$-sliced domain wall in comparison with those of the two AdS$_7$ vacua.
	}
	\label{fig:interpolatingDW}
\end{figure}
 Given the numerical solution for the superpotential, we can finally integrate the first-order equations \eqref{1storderflow}. For the sake of clarity we rewrite them here,
\begin{equation}\label{1storderflow1}
  A'=\frac{1}{60}\,e^{-6 A}\partial_A\,F\qquad \text{and} \qquad X'=-\frac{1}{10}\,e^{-6 A}X^2\,\partial_X\,F\,.
\end{equation}
We perform the numerical shooting using as initial value the non-SUSY vacuum placed at $r=0$. The result is plotted in Figure \ref{fig:interpolatingDW}. We thus obtained a domain wall solution interpolating between the non-supersymmetric and supersymmetric AdS$_7$ vacua.

\section{Bubble geometries in massive IIA}
\label{bubbles6d}

In this section we apply the same strategy of Section \ref{bubbles7d} to the case of smooth dS$_5$ domain walls in massive IIA supergravity. In analogy to dS$_6$ bubbles in M-theory we will work within the consistent truncation of massive IIA supergravity on a squashed 4-sphere constructed in \cite{Cvetic:1999un}. This truncation reproduces a very similar framework to the 7d supergravity used in Section \ref{bubbles7d}, especially if one considers only the scalar sector. In fact the lower-dimensional theory is the minimal incarnation of half-maximal gauged supergravity in 6d and admits two AdS$_6$ vacua associated with two different values of a single scalar field. In analogy with the 7d case, one vacuum is supersymmetric, while the other is not.

In this section we will adopt the Hamilton-Jacobi formulation of classical dynamical systems in this 6d setup with the purpose of searching for smooth dS$_5$ bubbles connecting the above two different vacua. To this aim we will apply the perturbative technique of Section \ref{perturbativeHJ} to this particular framework.

\subsection{Romans supergravity}\label{6dtheory}

Let us recall the main features of the consistent truncation of massive IIA supergravity constructed in \cite{Cvetic:1999un}. This is a warped compactification defined by an internal squashed 4-sphere and retaining only the fields belonging to the 6d supergravity multiplet without any matter multiplet. The isometry group is $\mathbb{R}^+\times \mathrm{SO}(4)$ and preserves 16 real supergracharges, namely it is $\ma N=(1,1)$ theory in 6d. In addition to 6d gravity, the field content is given by one real scalar field $X$, three SU$(2)$ vectors $A^i$, one abelian vector $A^0$ and a 2-form $B_{(2)}$. This 6d gauged supergravity is usually called Romans supergravity \cite{Romans:1985tw}.
We are interested in the case where all the vectors and the 2-form are vanishing. 
The Ansatz for the metric and dilaton takes the form \cite{Cvetic:1999un}
\begin{equation}
 \begin{split}\label{ansatzmetric6d}
  ds^2_{10}&=X^{-1/2}\,\Delta^{1/2}\,s^{-1/3}\,\left[ds^2_6+2g^{-2}\,X^{2}\,ds_{4}^2\right]\ ,\\
  ds_{4}^2&=d\xi^2+\Delta^{-1}\,X^{-3}\,c^{2}\,ds^2_{S^3}\ ,\qquad \text{with}\qquad \Delta=X\,c^2+X^{-3}\,s^2\ ,\\
  e^{\Phi}&=s^{-5/6}\,\Delta^{1/4}\,X^{-5/4}\ ,
 \end{split}
\end{equation}
 and $s=\sin \xi$ and $c=\cos\xi$. After imposing that the vectors and the 2-form are vanishing, only the 4-form flux and the Romans' mass $F_{(0)}=m$ survive, namely \cite{Cvetic:1999un}
\begin{equation}
\label{10dfluxesto6d}
 \begin{split}
  F_{(4)}&=-\frac{4\sqrt 2}{3}\,g^{-3}\,\Delta^{-2}\,\left(X^{-6}\,s^2-3X^2\,c^2+4\,X^{-2}\,c^2-6\,X^{-2} \right)\,s^{1/3}\,c^3\,d\xi\,\wedge\,\text{vol}_{S^3}\\
  &-8\sqrt{2}\,g^{-3}\,\Delta^{-2}\,X^{-3}\,\,s^{4/3}\,c^4\,dX\,\wedge\,\text{vol}_{S^3}\ .
 \end{split}
\end{equation}
The deformations of the 6d theory produced by this truncation are defined by two embedding tensor parameters, $g$ and $m$. As in the 7d case the first one is associated to the gauged R-symmetry group SU$(2)_R$ and the second is a St\"uckelberg mass for the 2-form. The scalar potential has the form \cite{Romans:1985tw,Cvetic:1999un}
\begin{equation}\label{6dpot}
 V=m^2 X^{-6}-12 m^2\,X^{-2}-9\,m^{2}\,X^2\ ,
\end{equation}
where we fixed $g=\frac{3m}{\sqrt 2}$ as required by the truncation Ansatz of \cite{Cvetic:1999un}. The 6d Lagrangian has the form
\begin{equation}
\label{6dlagrangian}
 \sqrt{-g}^{,.-1}\,\ma L=  R-4 \,X^{-2}\, \partial_\mu X\,\partial^\mu X-V\ .
\end{equation}
By taking the variation with respect to the metric and the scalar $X$ one can easily obtain the equations of motion,
\begin{equation}
 \begin{split}\label{eom6d}
  R_{\mu\nu}-4\,X^{-2}\,\partial_\mu X \partial_\nu X-\frac 14 \, V\, g_{\mu\nu}&=0\ ,\\
  \partial_\mu \left(\sqrt{-g}\,X^{-1}\,g^{\mu\nu}\,\partial_\nu X\right)-\frac{\sqrt{-g}}{8}\,X\,\partial_X V&=0\ .
 \end{split}
\end{equation}
This theory admits two different AdS$_6$ vacua, one supersymmetric and one not. Let us now consider them separately as in the case of 7d supergravity studied in Section \ref{bubbles7d}.

\vspace{.5cm} {\bf SUSY AdS$_6$ vacuum: $X=1$}

\noindent This is the Brandhuber and Oz vacuum describing the near-horizon geometry of the D4-D8 branes \cite{Brandhuber:1999np}. It is realized for $X=1$ and preserves 16 real supercharges. The 10d geometry is defined by a warped product of AdS$_6$ with a 4-sphere\footnote{More precisely the internal manifold is the upper hemisphere of a $S^4$ written as a foliation of $S^3$ over a segment parametrized by the coordinate $\xi$.} $S^4$,
\begin{equation}
 \begin{split}\label{SUSYAdS6}
  ds^2_{10}&=s^{-1/3}\left[ds^2_{\text{AdS}_6}+2g^{-2}\,ds_{S^4}^2\right]\ ,\\
 e^{\Phi}&=s^{-5/6}\ ,\\
  F_{(4)}&=\frac{20\sqrt 2}{3}\,g^{-3}\,s^{1/3}\,c^3\,d\xi\,\wedge\,\text{vol}_{S^3}\ .
 \end{split}
\end{equation}
with the radius of AdS$_6$ given by $L_{\text{SUSY}}=\frac{3}{\sqrt 2\,g}=m^{-1}$.

\vspace{.5cm} {\bf Non-SUSY AdS$_6$ vacuum: $X=3^{-1/4}$}

\noindent The non-supersymmetric AdS$_6$ vacuum of Romans supergravity is defined by $X=3^{-1/4}$. The geometry of this vacuum takes the following form,
\begin{equation}
\begin{split}\label{nonSUSYAdS6}
   ds^2_{10}&=s^{-1/3}\,[\,(3-2\,c^2)^{1/2}\,ds^2_{\text{AdS}_6}+2g^{-2}3^{-1/2}(3-2\,c^2)^{1/2}\,d\xi^2 +2g^{-2}3^{1/2}(3-2\,c^2)^{-1/2}c^2\,d^2_{S^3} ]\ ,\\
  e^{\Phi}&=3^{1/4}\,s^{-5/6}\,(3-2\,c^2)^{1/4}\ ,\\
  F_{(4)}&=12 \sqrt 2\,g^{-3}\,(3-2\,c^2)^{-2}\,s^{1/3}\,c^3\,d\xi\,\wedge\,\text{vol}_{S^3}\ .
 \end{split}
\end{equation}
with the radius of AdS$_6$ given by $L_{\slashed{\text{SUSY}}}=\frac{3^{1/4}\,5^{1/2}}{2^{1/2}\,g}=\frac{\sqrt 5}{3^{3/4}m}$.
As for the non-supersymmetric AdS$_7$ \eqref{nonSUSYAdS7}, the brane origin of this vacuum is not clear.

\subsection{First-order formulation for dS$_5$ domain walls}

Let us focus on domain wall geometries of the following form
\begin{equation}
 \begin{split}\label{dSDW6d}
  ds^2_{6}&=e^{2A(r)}\,L^2\,ds^2_{\text{dS}_5}+dr^2\ ,\\
  X&=X(r)\ .
 \end{split}
\end{equation}
The equations of motion \eqref{eom6d} take the form of two ODEs and the hamiltonian constraint,
\begin{equation}
 \begin{split}\label{eomDW6d}
   A''+5 A'^{\,2}-\frac{4\,e^{-2A}}{L^2}+\frac{1}{4}\, V&=0\ ,\\
    X''+5 A'\,X'-\frac{X'^{\,2}}{X^{2}}-\frac{1}{8}\,X^2\, \partial_X\,V&=0\ ,\\
    20A'^{\,2}+\frac{4\,X'^{\,2}}{X^{2}}-\frac{20\,e^{-2A}}{L^2}+V&=0\ ,
 \end{split}
\end{equation}
where the derivative with respect to $r$ has been denoted by $'$.
The geometry of AdS$_6$ vacua can be recovered by choosing
\begin{equation}\label{Avacua6d}
 A=\log \left(\sinh\left(\frac{r}{L} \right)\right)\ ,
\end{equation}
and imposing $X=1$ and $L_{\text{SUSY}}=m^{-1}$ or $X=3^{-1/4}$ and $L_{\slashed{\text{SUSY}}}=\frac{\sqrt 5}{3^{3/4}\,m}$ respectively for supersymmetric and non-supersymmetric vacua discussed in the previous section.

Let's construct the quantities needed in order to cast the second-order problem \eqref{eomDW6d} in a system of first-order ODEs. We can follow the same strategy of Section \ref{HJ7d} and outlined in general in Appendix \ref{HJmethod}. It is easy to show that the second-order equations for $A$ and $X$ written in \eqref{eomDW6d} can be obtained by taking the variation of the following 1d effective Lagrangian
\begin{equation}\label{effectiveL6d}
 L_{\text{eff}}=20\,e^{5A}\,A'^{\,2}-4\,e^{5A}\,\frac{\,X'^{2}}{X^{2}}+\frac{20\,e^{3A}}{L^2}-e^{5A}\,V\,.
\end{equation}
From the expression \eqref{classicalH} we can derive the conjugate momenta $\pi_A=40\,e^{5A}\,A'$ and $\pi_X=-8\,e^{5A}\,X^{-2}\,X'$. Taking the Legendre transformation of \eqref{effectiveL6d} we easily obtain the corresponding Hamiltonian
\begin{equation}\label{Heff6d}
 H_{\text{eff}}=\frac{1}{80}\,e^{-5A}\,\pi_A^{2}-\frac{1}{16}\,e^{-5A}\,X^{2}\,\pi_X^{\,2}-\frac{20\,e^{3A}}{L^2}+e^{5A}\,V\,.
\end{equation}
We are now ready to write the Hamilton-Jacobi equation for the superpotential $F=F(A,X)$ associated to the system \eqref{eomDW6d}. By expressing the momenta as $\pi_A=\partial_A\,F$ and $\pi_X=\partial_X\,F$, one obtains
\begin{equation}\label{HJequation6d}
 \frac{1}{80}\,e^{-5A}\,(\partial_A\,F)^{2}-\frac{1}{16}\,e^{-5A}\,X^{2}\,(\partial_X\,F)^{2}-\frac{20\,e^{3A}}{L^2}+e^{5A}\,V=0\,,
\end{equation}
where the constant $E$ appearing in \eqref{genericHJequation} must be zero in order to satisfy the Hamiltonian constraint. As we did in Section \ref{HJ7d} we reduced the the second-order problem \eqref{eomDW6d} to a single PDE. This equation need to be satified by a suitable solution for the superpotential $F$. Once the solution of the Hamilton-Jacobi equation is found, the radial flow featuring the domain wall \eqref{dSDW6d} can be worked out easily through by intergrating the first-order equations \eqref{general1STorder}. In this specific case they have the form
\begin{equation}\label{1storderflow6d}
  A'=\frac{1}{40}\,e^{-5 A}\partial_A\,F\qquad \text{and} \qquad X'=-\frac{1}{8}\,e^{-5 A}X^2\,\partial_X\,F\,.
\end{equation}
The solutions of \eqref{1storderflow6d} solve automatically also the equations of motion since, as explained in Appendix \ref{HJmethod}, they imply the extremization of the action \eqref{effectiveL6d}.

\subsection{The interpolating solution}

In this section we will follow the same strategy of numerical integration of Section \ref{strategies}. The idea is to solve perturbatively the Hamilton-Jacobi equation \eqref{HJequation6d} by expanding around the non-supersymmetric AdS$_6$ vacuum \eqref{nonSUSYAdS6}. Let us start by checking the reliability of the perturbative analysis. To this aim we may compare the profile of the scalar potential $V(X)$ given in \eqref{6dpot} with its Taylor expansion $V_{\text{pert}}(X)$ around the non-supersymmetric vacuum $X=3^{-1/4}$.
\begin{figure}[!http]
	\centering
	\includegraphics[width=0.75\textwidth]{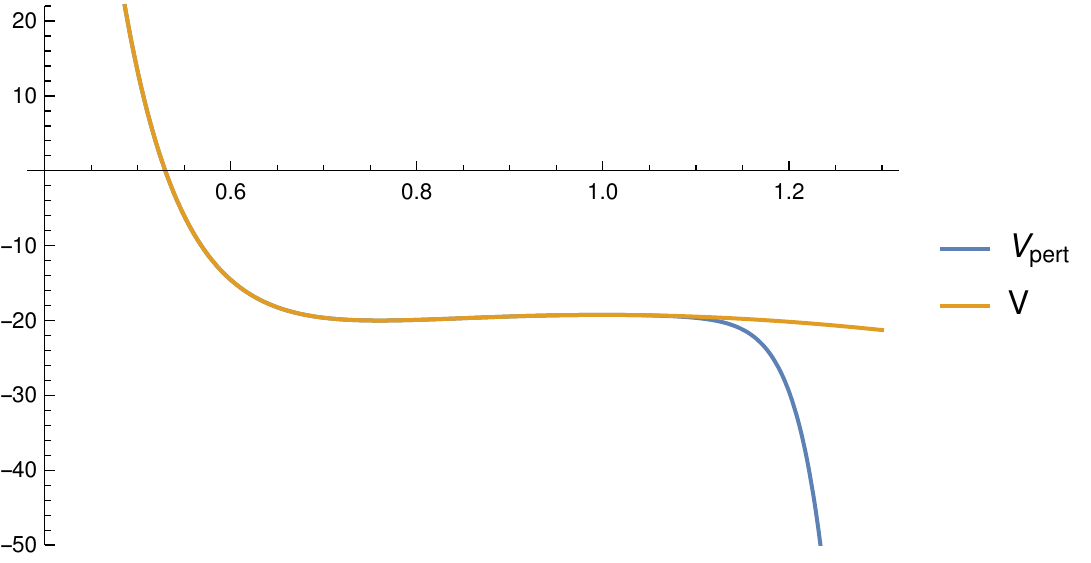}
	\caption{\it  The comparison between the scalar potential $V(X)$ given in \eqref{6dpot} and its expansion at 15th order around the non-SUSY critical point $X=3^{-1/4}\simeq 0.75$. For values of $X$ between the two critical points $X\,\in\,[3^{-1/4},\,1]$, the scalar potential is well-approximated by its perturbative expansion. The plot is given for $m=\frac{\sqrt{2}}{3}\,g=1$.
	}
	\label{fig:comparePotentials6d}
\end{figure}
From Figure \ref{fig:comparePotentials6d} it is manifest that if the perturbative order is sufficiently high then the scalar potential can be substituted by its perturbative expansion in the interval $X\,\in\,[3^{-1/4},\,1]$.

In order to solve the Hamilton-Jacobi equation we need to impose suitable boundary conditions on the derivatives of the fake superpotential. It is easy to verify that the expression 
\begin{equation}\label{Finfty6d}
 F_{\infty}=\frac{10}{L}\,e^{4A}-\frac{L}{6}\,e^{6A}\,V
\end{equation}
solves the equation \eqref{HJequation6d} in small $L$ limit, up to orders $O(L^2)$. The expression $F_{\infty}$ describes the highly-curved regime of small values of $L$ in which the contribution of dS$_5$ is dominant in the stress-energy tensor. We can use \eqref{Finfty6d} to produce the initial conditions on the derivatives of $F(A,X)$ at each order of the perturbative expansion.
Let us recall the Ansatz \eqref{perturbativeF} for the fake superpotential expanded around the non-supersymmetric vacuum
\begin{equation}
 F(A,X)=4\,\sum_{k=0}^{\infty}F^{(k)}(A)\,\frac{(X-3^{-1/4})^k}{k!}\ ,
\end{equation}
where the coefficients $F^{(k)}$ need to depend on $A$ to take into account the non-separability of the effective potential defining the Hamiltonian \eqref{Heff6d}. With this Ansatz, the Hamilton-Jacobi equation \eqref{HJequation6d} becomes a set of ODEs for the coefficients $F^{(k)}(A)$, each for any order in $k$. 
As for the 7d case we impose $F^{(1)}(A)=0$ in order to keep each perturbative order decoupled. At the 0-th order one obtains
\begin{equation}\label{zeroorder6d}
 \frac{1}{5}\,e^{-5A}\,(\dot F^{(0)})^2-\frac{20\,e^{5A}}{L^2}-\frac{20\,e^{3A}}{L^2}=0\ ,
\end{equation}
\begin{figure}[!http]
\centering
	\includegraphics[scale=0.5]{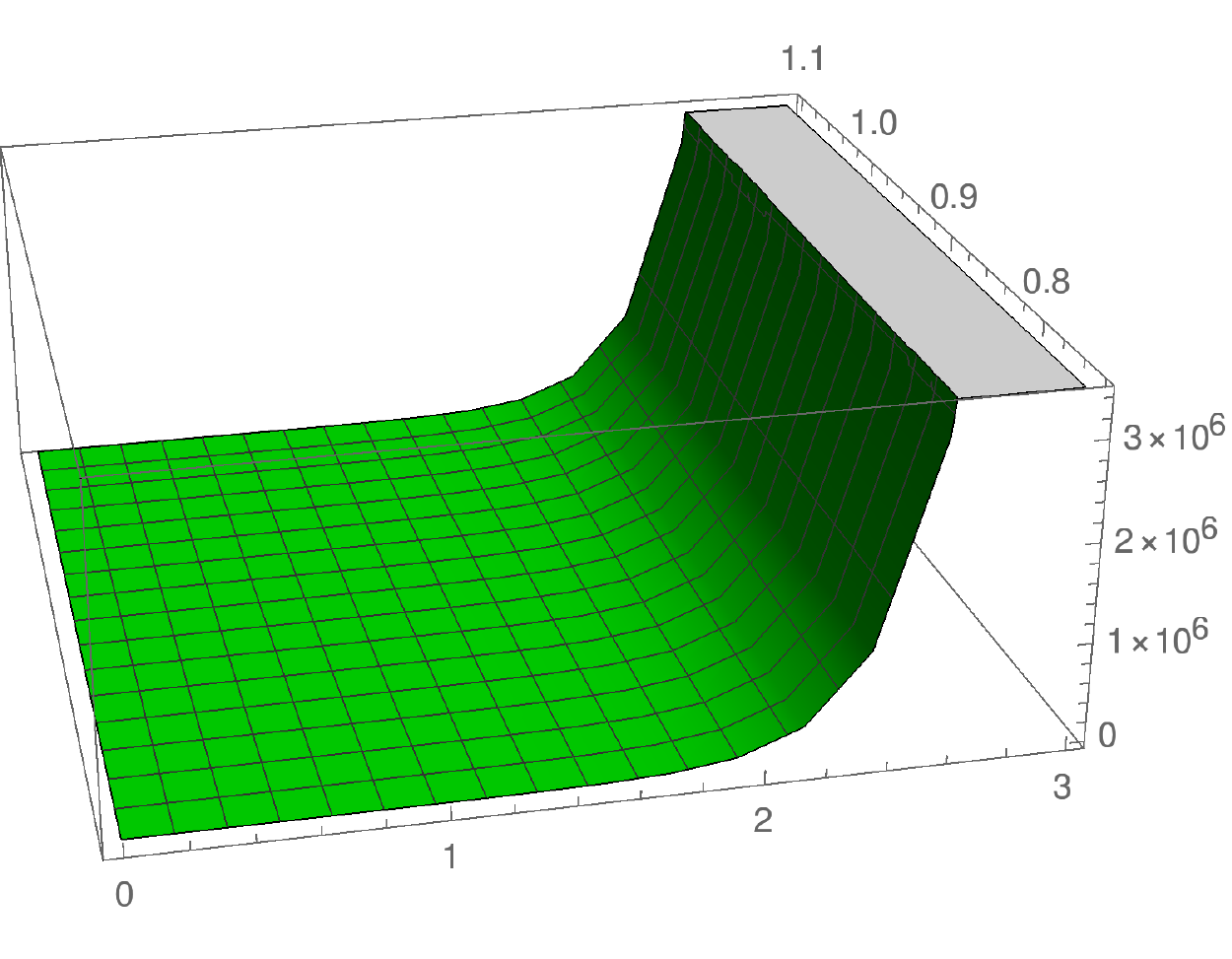}\qquad\quad \,\,\,
	\centering
	\includegraphics[scale=0.6]{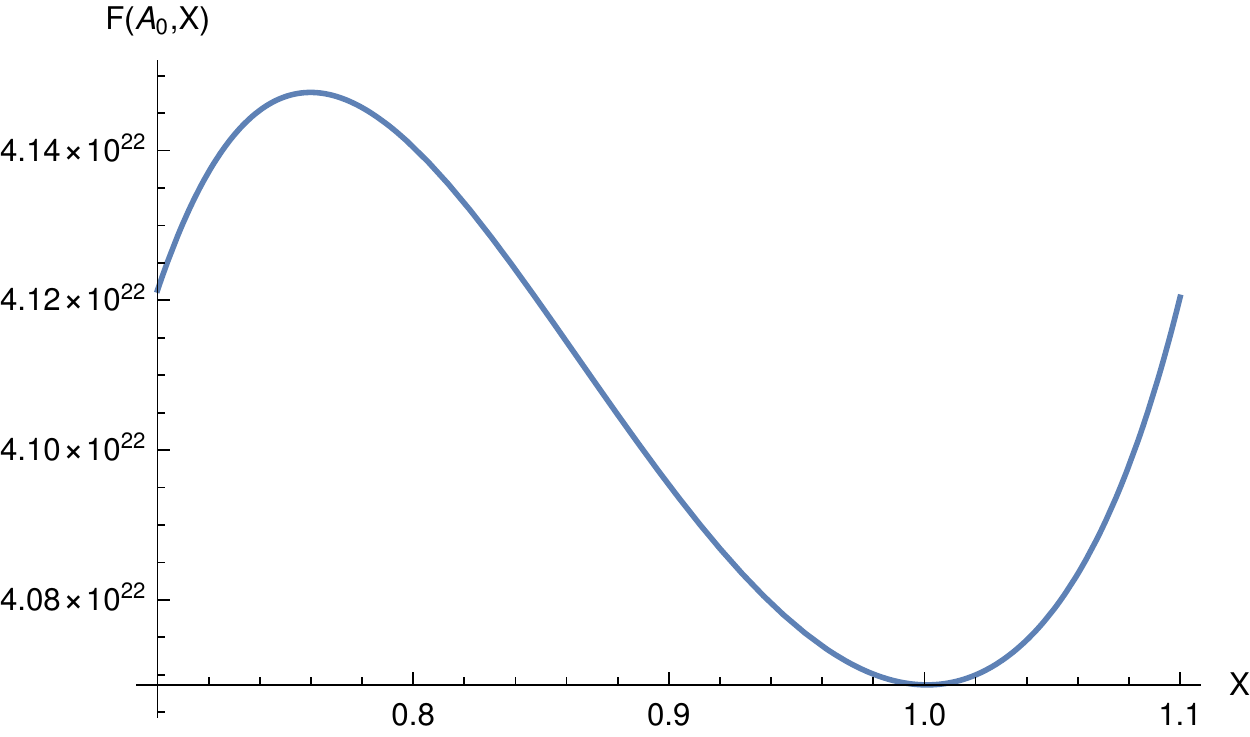}
	\caption{\it The profile of the superpotential $F(A,X)$ obtained with perturbative intergration up at order $k=15$. The 3d plot describes the full solution in the intervals $A\in[0,\,3]$ and $X\in[0.74,1.1]$. On the right side the same function is plotted for $A_0=10$. Even if the variation of $F$ is tiny, we observe an interpolating profile between the two AdS$_6$ vacua respectively located at $X=3^{-1/4}\simeq0.75$ and $X=1$. The plots are given for $L=1$.
	}
	\label{fig:plotSuperpotential6d}
\end{figure}
where the derivative with respect to $A$ has been denoted by $\cdot$ and we expressed the scalar potential in terms of the radius of non-SUSY vacuum through the relation $m=\frac{3^{3/4}}{\sqrt 5}L$ characterizing the non-SUSY vacuum \eqref{nonSUSYAdS6}. This equation can be solved exactly, this procedure leading to an expression similar to \eqref{Fzero}. After determining the profile of $F^{(2)}(A)$ numerically, one can integrate each ODE belonging to the perturbative tower finding the coefficients $F^{(k)}(A)$.
At each order in $k\geq2$ the numerical integration one has to evaluate the $(k-1)$-th solution of the previous step in the $k$-th ODE for $F^{(k)}(A)$. As we did for the initial conditions \eqref{initF} for the 7d case, we impose the following boundary conditions
\begin{equation}
 F^{(k)}=4^{-1}\,\partial^k_{X}\,F_{\infty}^{(k)}|_{X=3^{-1/4}}\ ,
\end{equation}
where $F_{\infty}^{(k)}$ is given by \eqref{Finfty6d} and it has been written in terms of the perturbative expansion of the scalar potential $V$ at the $k$-th order.
The solution for $F(A,X)$ obtained by iterating the numerical integrations up to the order $k=15$ is plotted in Figure \ref{fig:plotSuperpotential6d}. Also in this case we observe an interpolating flow of the superpotential between the two vacua.
\begin{figure}[!http]
\centering
	\includegraphics[scale=1]{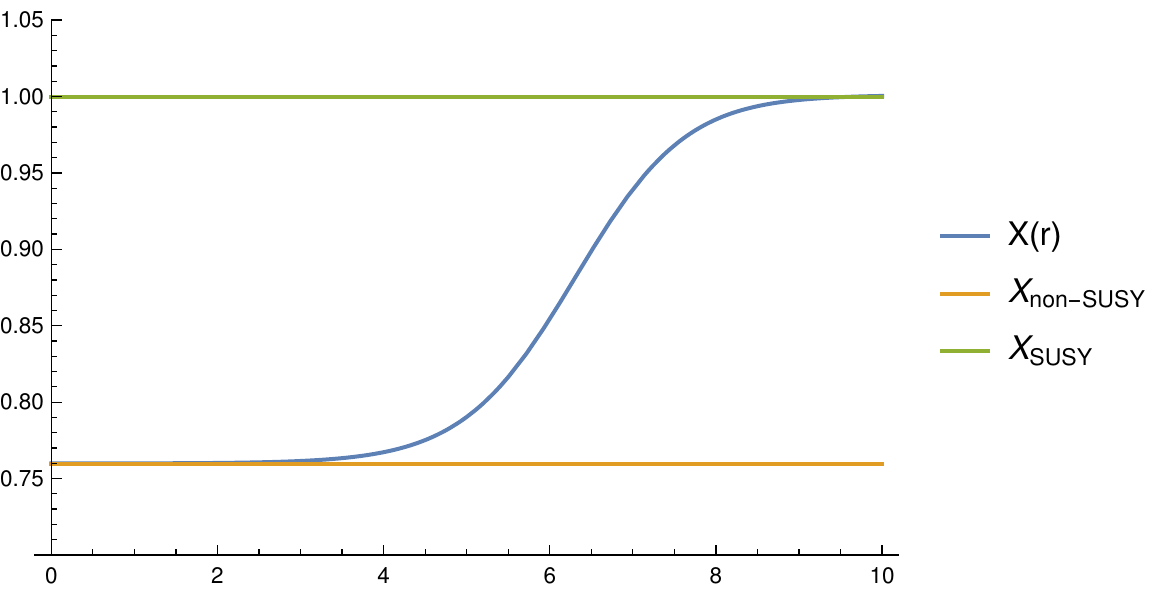}
	\caption{\it The smooth radial flows of the scalar $X$ of the dS$_5$-sliced domain wall in comparison with those of the two AdS$_6$ vacua.
	}
	\label{fig:interpolatingDW6d}
\end{figure}
The last step is integrating the first-order equations \eqref{1storderflow6d} using the numerical solution for $F(A,X)$. We perform this radial intergration by taking as initial value the non-supersymmetric vacuum placed at $r=0$. The result of this integration for $A$ and $X$ is plotted in Figure \ref{fig:interpolatingDW6d}. We observe that the radial flow of the domain wall smoothly connect the non-supersymmetric AdS$_6$ vacuum to the Brandhuber-Oz vacuum of massive IIA.

\section{Final comments}
In this work we presented the methology to provide the geometries describing gravitational instantons connecting two different AdS vacua, one preserving SUSY and the other not. The key feature of our solutions is that they are fully backreacted in the sense that we did not have to make use of probe calculations in order to study a tunneling process, nor have we imposed the thin-wall approximation at the interface between the two vacua.

Nevertheless there are many features of our solutions that still require a deeper understanding and this paper aims at representing a first intermediate step in this research direction. A quite misterious element is related to the brane interpretation of our 7d and 6d dS domain walls. Interesting results regarding the microscopic origin of the non-pertubative instability of the non-SUSY AdS vacua considered in this paper have been obtained by looking at the 7d supergravity as a compactification of massive IIA string theory \cite{Apruzzi:2016rny,Apruzzi:2019ecr}. For what concerns the 6d case, in \cite{Apruzzi:2021nle} the analysis on the non-perturbative instability of non-SUSY AdS$_6$ vacuum has been performed by embedding the Romans' supergravity in Type IIB. An ambitious step forward would be obtaining the aforementioned non-SUSY vacuum geometries as the near-horizon regime of a suitable brane solution or, at least, to provide a clear understanding of the microscopic objects underlying their non-perturbative decays in the spirit of the analysis of the (singular) solutions of \cite{Horowitz:2007pr} and, more recently, of \cite{Bomans:2021ara}.

Furthermore, a relevant aspect to investigate is the possibility of applying our solutions to string cosmology. In \cite{Banerjee:2018qey,Banerjee:2019fzz} a proposal of dS$_4$ cosmology was formulated in the context of Type IIB starting from the imposition of Israel junction conditions between two AdS$_5$ vacua, one of which is supersymmetric and the other not. The mechanism identified in the aforementioned references as the responsable of the emergence of de Sitter geometry at the interface of the two AdS vacua was exactly the Coleman-de Luccia decay of the vacuum with broken SUSY. With the approach outlined in this paper it would be interesting to test this proposal by deriving the fully backreacted solution describing this construction in Type IIB.

Another interesting issue that seems to require further clarifications and better understanding is the relation of our work to positive energy theorems in string compactifications. Strictly speaking, the existence of our dS domain walls may appear in contradiction with the analysis in \cite{Danielsson:2016rmq,Dibitetto:2022vdw}, where the existence of global fake superpotentials bounding the scalar potential from below is argued to be a sufficient condition for vacuum stability.
Our present findings seem to restrict the validity of such positive energy theorems in a gauged supergravity to the set of solutions corresponding to a given choice of branch for the superpotential. Other flows associated with different fake superpotentials belong to disconnected subsets of the space of solutions to the Hamilton-Jacobi equations. Due to this, globally bounding superpotentials might after all have nothing conclusive to say about non-perturbative (in)stabilities. A decisive analysis one needs to go through in order to assess the contribution of our bubbles to the Euclidean path integral is to extract the effective wall tension and see whether it respects the CDL bound. However, we should also stress that this procedure for thick walls is not free of subtleties and ambiguities. We hope to come back to this issue in the fututre.

We would like to conclude by reflecting on a final aspect concerning our present results. Back in the original literature from the 1980's, a semiclassical Euclidean path integral approach was used, in order to come up with physical predictions for false vacuum decay. We should definitely stress that our 6d \& 7d theories arising from consistent truncations of string and M-theory can never be regarded as good effective theories in a Wilsonian sense. We only used the trucntion formulas as a technique to produce 10d \& 11d solutions involving expanding bubbles within AdS. Hence, a proper estimation of the nucleation probability should be directly performed within the higher-dimensional description. 

\section*{Acknowledgements}

The work of NP is supported by the Israel Science Foundation (grant No. 741/20) and by the German Research Foundation through a German-Israeli Project Cooperation (DIP) grant ``Holography and the Swampland".

\appendix

\section{Hamilton-Jacobi method for classical systems}
\label{HJmethod}

In this appendix we schematically review the Hamilton-Jabobi (HJ) formulation in studying the dynamics classical systems\footnote{See also Appendices of \cite{Dibitetto:2020csn}, and \cite{Dibitetto:2022vdw} for more details.}. The power of this method is to allow to cast the equations of motion of a classical system in terms of a set of first order constraints, even in the absence of supersymmetry.

The starting idea is to formulate the variational principle for a given configuration by recasting the action into a sum of various squares. Requiring that each of these squares vanish separately, determines a set of first-order ODEs. The solutions of these first-order equations solve the equations of motion by construction.

The general form of these first-order conditions can be determined in generality as follows. Let us start by a generic action of the form
\begin{equation}
  S(q,r)=\int{dr}\,L\qquad \text{with}\qquad L=\frac12 M_{\alpha\beta}\dot q^\alpha \dot q^\beta-V(q)\ .
  \label{genericaction}
\end{equation}
The ``time'' parameter has been called $r$ since for the cases we are interested in, the aforementioned action is obtained by reducting some (super)gravity system to one dimension. Then the 1d ``time" typically describes the radial flow of some (super)gravity solution. The variables $q^\alpha(r)$ turn out to describe the dynamical functions associated to the gravity background under study (\emph{e.g.} the warp factors, scalar fields, etc). With a Legendre transformation one can derive the Hamiltonian.
\begin{equation}\label{classicalH}
 H=\frac12\,M^{\alpha\beta}p_\alpha p_\beta+V(q)\qquad \text{with}\qquad p_\alpha=\partial_{\dot q^\alpha}L=M_{\alpha\beta}\dot q^\beta\ .
\end{equation}

The core of this approach consists in the introduction of a ``superpotential'' $F(q)$. This function includes all the information to identify the dynamics of the system. The superpotential can be defined through the so-called Hamilton-Jacobi equation,
\begin{equation}
  H(\partial_q F, q)+\frac{\partial S}{\partial r}=\frac12 M^{\alpha\beta}\partial_\alpha F\,\partial_\beta F+V-E=0\qquad \text{with} \qquad p_\alpha=\partial_\beta F\ ,
   \label{genericHJequation}
\end{equation}
where the action has been crucially interpreted as a function of the dynamical variables with the form $S(q)=F(q)-r E$ with $E$ constant. We can now use \eqref{genericHJequation} to cast the action into a sum of squares. Expressing the potential $V(q)$ in terms of the superpotential by using \eqref{genericHJequation} and plugging the expression in the action we get up to total derivatives,
\begin{equation}
 S=\frac12\,\int{dr \,M_{\alpha\beta}\,\left(\dot q^\alpha-M^{\alpha\gamma}\partial_\gamma F\right)\left(\dot q^\beta-M^{\beta\delta}\partial_\delta F\right)}\ .
\end{equation}
Setting to zero each of these squared produce the following system of ODEs,
\begin{equation}\label{general1STorder}
 \dot q^\alpha=M^{\alpha\beta}\partial_\beta F\ .
\end{equation}
By construction the solutions of the above differential conditions extremize the action. As we mentioned at the beginning of this appendix this method does not rely on any supersymmetry completion when it is applied on a given supergravity background and, for this reason, it constitutes a good method in searching for non-supersymmetric solutions characterized by non-trivial radial flows.

	\bibliographystyle{utphys}
	\bibliography{references}
\end{document}